# Analytical and Experimental Study of a Variable-Mass Oscillator with Constant Mass-Loss Rate

*Rod Milbrandt*[1,*], *Josep Ll. Suñer*[2], *Juan C. Castro-Palacio*[3],
*Francisco M. Muñoz-Peréz*[3], *Juan A. Monsoriu*[3]

[1]Faculty of Physics and Engineering, Rochester Community and Technical College, Rochester, Minnesota, USA

[2]Instituto de Ingeniería Mecánica y Biomecánica, Universitat Politècnica de València, 46022 València, Spain

[3]Centro de Tecnologías Físicas, Universitat Politècnica de València, 46022 Valencia, Spain

*Corresponding author: rod.milbrandt@rctc.edu

## Abstract

We present an analytical and experimental study of a variable-mass oscillator with constant mass-loss rate. Starting from the equation of motion of a spring–mass system whose mass decreases linearly with time, analytical solutions are obtained in terms of Bessel functions. Simple asymptotic expressions are then derived, providing explicit descriptions of the evolution of the oscillation amplitude and frequency. These expressions provide a theoretical model suitable for direct comparison with experimental observations. Experimental measurements of the oscillator acceleration were performed using a simple laboratory setup consisting of a spring-mounted container that continuously loses mass at an approximately constant rate. The theoretical predictions accurately reproduce the measured dynamics over the entire duration of the oscillation, and the mass-loss rates obtained from the fits are in good agreement with independent measurements. The results provide an accessible example of variable-mass dynamics and illustrate how special-function methods can yield experimentally testable predictions in advanced undergraduate mechanics.

## 1. Introduction

Variable-mass systems constitute a classical topic in mechanics and provide instructive examples of open systems in which matter crosses the system boundary. Unlike constant-mass systems, their analysis requires a careful treatment of momentum transport and the application of Newton's second law, making them particularly valuable in advanced undergraduate mechanics.[1,2] Classical examples include rocket propulsion and other mass-flow systems,[1–3] as well as experimentally accessible configurations such as pulley-driven variable-mass systems.[4,5] Other widely studied examples involve the dynamics of chains in variable-mass configurations[6–9] and rope systems in which mass is continuously transferred between interacting subsystems.[10,11] These systems have attracted sustained interest from both theoretical and experimental perspectives because they illustrate the subtle interplay between external forces, momentum flux, and mass redistribution while remaining sufficiently accessible for laboratory investigations. More generally, variable-mass systems provide a useful framework for examining the fundamental principles of mechanics in open systems and the proper application of Newton's second law when mass is exchanged with the surroundings.[12]

Among the different classes of variable-mass systems, oscillators with continuously varying mass constitute a particularly interesting case because the changing mass affects both the oscillation amplitude and the frequency. One of the most extensively studied examples consists of a spring–mass oscillator in which the oscillating mass decreases continuously as granular material escapes from a container attached to the spring.[13] Such systems exhibit dynamics that differ significantly from those of the conventional harmonic oscillator, including a progressive increase in oscillation frequency as the mass decreases.[13] Additional theoretical studies have explored alternative formulations of variable-mass oscillators and have obtained analytical solutions for specific time-dependent mass laws, further highlighting the mathematical richness of these systems.[14] Related investigations have also considered oscillatory systems with variable mass and time-dependent frequencies from a signal-analysis perspective, emphasizing the non-stationary nature of the resulting motion.[15]

Despite the considerable attention devoted to variable-mass oscillators, obtaining analytical descriptions of their dynamics remains a challenging task because the governing equations generally involve time-dependent coefficients. Previous studies have generally relied on approximate treatments of the oscillatory motion,[13] while exact analytical solutions have been obtained only for particular forms of the mass variation law.[14] In the present work, the assumption of a constant mass-loss rate leads to a differential equation whose exact solution can be expressed in terms of Bessel functions. Beyond providing a closed-form analytical description of the dynamics, this formulation enables the derivation of simple asymptotic expressions in which the motion is represented by harmonic-like functions with slowly varying amplitude and frequency. These expressions provide a transparent physical interpretation of the oscillator dynamics and yield analytical predictions for the temporal evolution of the oscillation frequency, a quantity that can be directly extracted from experimental measurements.

To assess the validity of the proposed model, its predictions are compared with measurements obtained from a laboratory realization of the variable-mass oscillator. The experimental study combines independent determinations of the mass-loss rate from discharge measurements, nonlinear fits of the acceleration data, and an analysis of the time evolution of the oscillation frequency, providing several complementary tests of the theoretical predictions. The remainder of the paper is organized as follows. Section 2 presents the theoretical description of the variable-mass oscillator and derives the corresponding analytical solution. Section 3 describes the experimental setup and measurement procedure. Section 4 presents the comparison between theoretical predictions and experimental results, including the determination of the mass-loss rate for two discharge conditions and the experimental verification of the predicted frequency evolution. Finally, Section 5 summarizes the main conclusions.

# 2. Theoretical Analysis

The system considered in this work is a vertical variable-mass oscillator consisting of a container suspended from a spring of spring constant $k$, as shown in Fig. 1. The container is partially filled with granular material that continuously leaves the system through a small opening at its bottom. As a result, the total oscillating mass decreases with time while the spring constant remains unchanged. The gradual reduction of the mass modifies the dynamical response of the oscillator, leading to a time-dependent oscillation frequency that differs from that of a conventional harmonic oscillator with constant mass.

To simplify the analysis, we assume that the thrust associated with the momentum carried away by the escaping grains is negligible compared with the forces governing the oscillatory motion, following Ref. [5]. The mass flow is assumed to be uniform throughout the experiment. The model also assumes predominantly vertical motion and neglects dissipative effects. The contribution of the spring mass is accounted for through an effective oscillating mass, as described in Sec. 3. The analytical solution derived below is exact within these assumptions and is not intended as a general description of all granular-discharge oscillators.

For a constant mass-loss rate, the mass of the oscillating system decreases linearly with time according to

$$m(t) = m_0 - \alpha t \qquad (1)$$

where $m_o$ denotes the initial mass and $\alpha$ is the mass-loss rate. The parameter $\alpha$ can be written as

$$\alpha = -\frac{dm}{dt}. \qquad (2)$$

where the negative sign reflects the fact that the mass decreases with time. The displacement $x(t)$ is defined relative to the initial equilibrium position $x_0$, taking the downward direction as positive.

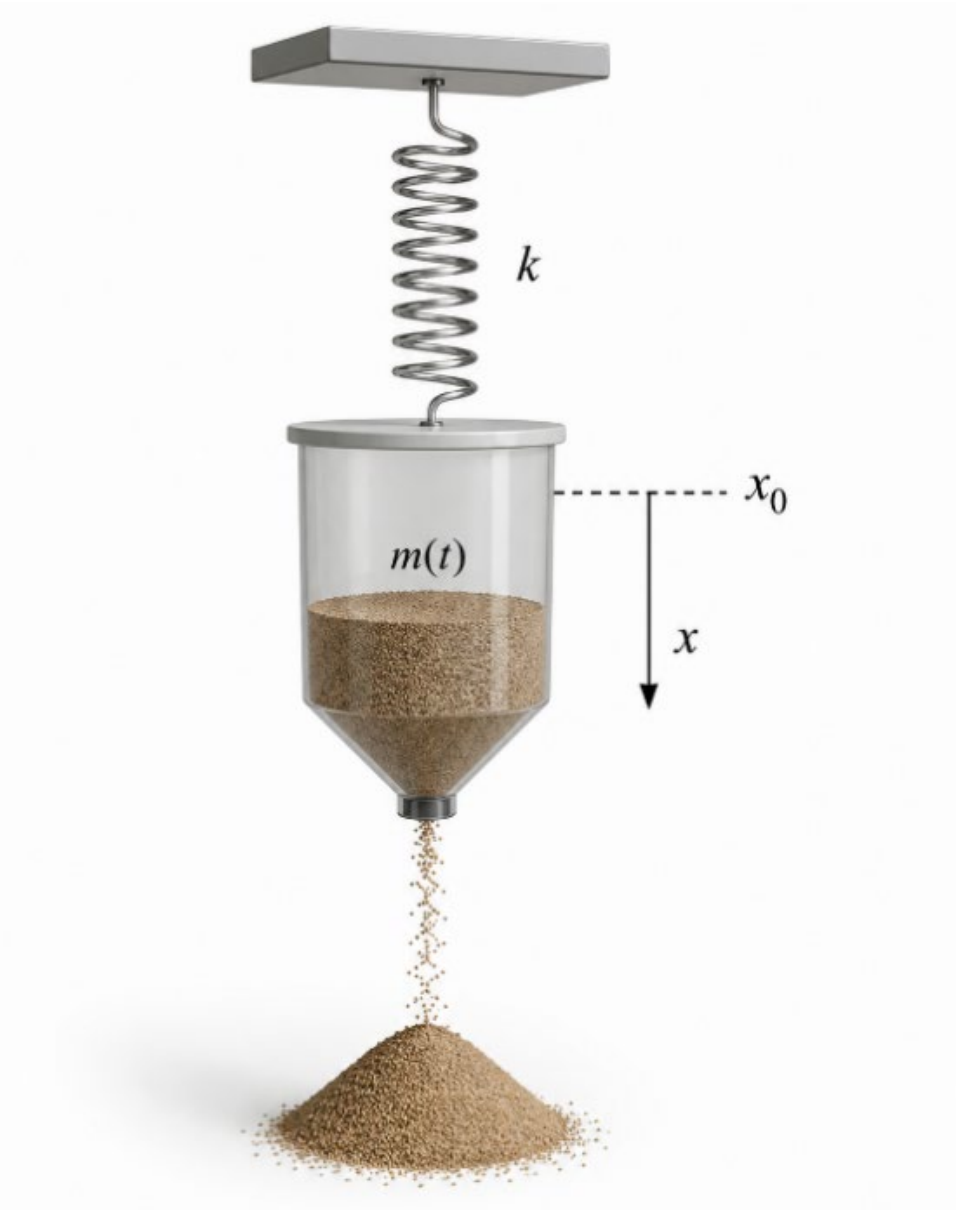

**Fig. 1.** Schematic representation of the variable-mass oscillator considered in this work. A container suspended from a spring of spring constant $k$ continuously loses granular material through a small opening at its bottom, resulting in a time-dependent mass $m(t)$. The displacement $x(t)$ is measured from the initial equilibrium position $x_0$, with the downward direction taken as positive.

## A. Analytical solution of the variable-mass oscillator

The motion of the variable-mass oscillator can be obtained by applying Newton's second law to the system shown in Fig. 1. The linear momentum of the oscillating mass is

$$p(t) = m(t)\dot{x} \tag{3}$$

The gravitational force acting on the system is $F_g = m(t)g$ and the elastic restoring force is $F_e = -k(x_0 + x)$, where $x_0$ is the initial equilibrium position, measured as the elongation of the spring from its unstretched length, and $x$ is the displacement measured from that fixed reference position. As the mass decreases, the instantaneous equilibrium position changes, while $x_0$ remains fixed.

For this open system, Newton's second law must account for the momentum carried away by the escaping material. Under the negligible-thrust approximation introduced above, the equation of motion becomes

$$m(t)g - k(x_0 + x) = m(t)\ddot{x}. \tag{4}$$

Using the equilibrium condition $m_o g = kx_0$, and substituting Eq. (1) for the time-dependent mass, we obtain

$$-\alpha t g - kx = (m_0 - \alpha t)\ddot{x}. \tag{5}$$

Equation (5) governs the displacement of the variable-mass oscillator. In principle, this equation may be solved directly to determine the position x(t) as a function of time and compared with experimental position measurements. Such an approach is commonly employed in instructional laboratories using video-analysis tools such as Tracker, which allow the motion of oscillatory systems to be recorded and quantitatively analyzed from digital video recordings. Similar methodologies have been successfully applied to oscillatory systems such as coupled oscillators and normal-mode analysis.[16]

An alternative strategy consists of monitoring the motion using dedicated sensors. In recent years, a wide variety of low-cost data-acquisition devices have become available for physics teaching and laboratory activities, including smartphone sensors, Arduino-based systems, BBC micro:bit devices, PocketLab platforms, wireless motion sensors, and other portable measurement tools. Comprehensive reviews of these technologies and their educational applications can be found in the literature.[17] These tools enable the direct measurement of physical quantities such as acceleration, angular velocity, magnetic field, pressure, sound intensity, and light levels, among many others.

In the present work, we choose to monitor the system through acceleration measurements obtained with the accelerometer of a smartphone. Smartphone accelerometers have previously been shown to provide accurate measurements in studies of free and damped harmonic oscillations.[18] In addition, differentiating the equation of motion twice eliminates the time-dependent inhomogeneous term associated with the drifting equilibrium position, yielding a homogeneous differential equation for the acceleration. Consequently, it is advantageous to reformulate the problem directly in terms of the acceleration. Differentiating Eq. (5) twice and introducing $a = \ddot{x}$ yields

$$(m_0 - \alpha t)\ddot{a} - 2\alpha\dot{a} + ka = 0 \tag{6}$$

The above equation constitutes the fundamental equation for the analysis that follows. To solve it, we introduce the dimensionless variable $z(t) = 2\sqrt{km(t)}/\alpha$, and express the acceleration as $a(t) = u[z(t)]/\sqrt{m(t)}$, where $u(z)$ is an auxiliary function. Substituting these expressions into Eq. (6) yields

$$z^2\frac{d^2u}{dz^2} + z\frac{du}{dz} + (z^2 - 1)u = 0 \tag{7}$$

which is the standard Bessel differential equation of order one. Its general solution is $u(z) = C_1 J_1[z] + C_2 Y_1[z]$ where $J_1(z)$ and $Y_1(z)$ are Bessel functions of the first and second kinds, respectively, and $C_1$ and $C_2$ are constants determined by the initial conditions. Therefore, the acceleration can be written as

$$a(t) = \frac{C_1 J_1[z(t)] + C_2 Y_1[z(t)]}{\sqrt{m(t)}} \tag{8}$$

with $z(t) = 2\sqrt{km(t)}/\alpha$ and $m(t) = m_o - \alpha t$.

Equation (8) provides an exact analytical expression for the acceleration of the variable-mass oscillator. While exact, the solution does not readily reveal how the oscillation amplitude and frequency evolve as the mass decreases. To obtain a more transparent physical interpretation, we examine the asymptotic behavior of the Bessel functions. For the experimental conditions considered here, the value of z ranges from 364 to 619 (small hole) and from 95 to 161 (large hole), so our use of the asymptotic regime is justified.

In the limit of large values of $z$, corresponding to sufficiently large masses or slow mass-loss rates, the Bessel functions may be approximated by

$$J_1(z) \approx \sqrt{\frac{2}{\pi z}} \cos\left[z - \frac{3\pi}{4}\right], \text{ and } Y_1(z) \approx \sqrt{\frac{2}{\pi z}} \sin\left[z - \frac{3\pi}{4}\right] \quad (9)$$

Substituting these expressions into Eq. (8) yields

$$a(t) \approx \sqrt{\frac{2}{m(t)\pi z(t)}} \left(C_1 \cos\left[z(t) - \frac{3\pi}{4}\right] + C_2 \sin\left[z(t) - \frac{3\pi}{4}\right]\right). \quad (10)$$

Using standard trigonometric identities, the acceleration can be written in the more compact form

$$a(t) \approx \frac{C_3}{\sqrt{m(t)z(t)}} \cos[z(t) + \phi] \quad (11)$$

where $C_3$ and $\phi$ are constants determined by the initial conditions.

Equation (11) shows that the acceleration retains an approximately harmonic form, but with a time-dependent amplitude and phase. Since $z(t) = 2\sqrt{km(t)}/\alpha$, the amplitude factor in Eq. (11) scales as $a_{\max} \propto m(t)^{-3/4}$. The asymptotic solution therefore predicts a gradual increase in acceleration amplitude as the mass decreases. At the same time, the argument of the cosine term depends on $z(t) \propto \sqrt{m(t)}$, indicating that the oscillation frequency also evolves continuously during the discharge process. Consequently, the system behaves as a non-stationary oscillator whose amplitude and frequency vary throughout the discharge process. This asymptotic description captures the essential features of the variable-mass oscillator within the assumptions of the proposed model.

For the two outlet diameters investigated experimentally (11.60 and 19.55 mm, as described in Sec. 3), the dimensionless parameter $z$ ranges from 364 to 619 and from 95 to 161, respectively. The leading correction to the asymptotic expansion of the first-order Bessel functions is proportional to $3/(8z)$ corresponding to less than 0.4% even at the smallest experimental value of $z$. These large values justify the use of the asymptotic approximation. Moreover, an important advantage of Eq. (11) over the exact Bessel-function solution, Eq. (8), is its clear physical interpretation: it describes an oscillation with both time-dependent amplitude and instantaneous frequency. This simple expression is particularly suitable for connecting the theoretical model with the experimental observations and highlighting the underlying physics in an educational context.

### B. Time-dependent oscillation frequency

The asymptotic solution obtained in Sec. 2.A shows that the oscillatory motion is governed by a total phase of the form $\Phi(t) = z(t) + \phi$. Using $z(t) = 2\sqrt{km(t)}/\alpha$ and $m(t) = m_0 - \alpha t$, the phase function becomes

$$\Phi(t) = \frac{2\sqrt{k(m_0-\alpha t)}}{\alpha} + \phi \tag{12}$$

The instantaneous angular frequency $\omega$ is obtained from the time derivative of the phase, $\omega = \left|\frac{d\Phi}{dt}\right|$, which yields $\omega(t) = \sqrt{k/m(t)}$. Therefore, the oscillation frequency $f = \frac{\omega}{2\pi}$ is

$$f(t) = \frac{1}{2\pi}\sqrt{\frac{k}{m_0-\alpha t}} \tag{13}$$

It is worth noting that in the limit of zero mass loss, $\alpha = 0$, Eq. (13) reduces to the well-known expression for the natural frequency of a simple mass–spring harmonic oscillator. The present result therefore generalizes the classical harmonic-oscillator frequency to the case of a system with continuously decreasing mass.

Equation (13) shows that the oscillation frequency increases as the mass decreases. Since the oscillation period is given by $T = 1/f$, an equivalent prediction is obtained by rearranging the above equation as

$$T^2 = \frac{4\pi^2}{k}(m_0 - \alpha t) \tag{14}$$

which predicts a linear decrease of the square of the oscillation period with time. This relationship provides a straightforward experimental test of the model and will be examined experimentally for the different discharge conditions considered in this work.

An interesting consequence of the asymptotic solution is that the displacement and acceleration amplitudes evolve in opposite directions. Since the instantaneous angular frequency satisfies $\omega^2(t) = k/m(t)$, the oscillatory displacement amplitude can be estimated from the acceleration amplitude as $A_x(t) \approx A_a(t)/\omega^2(t)$. Taking into account that $A_a(t) \propto m(t)^{-3/4}$, we obtain $A_x(t) \propto m(t)^{1/4}$. Thus, while the acceleration amplitude increases as the mass decreases, the displacement amplitude progressively decreases. This distinction provides an additional physical insight into the dynamics of the variable-mass oscillator.

# 3. Experimental Setup

The experimental apparatus is shown in Fig. 2. A smartphone mounted in a commercial phone holder is suspended from a vertical spring attached to a laboratory stand. The smartphone serves both as part of the oscillating system and as a data-acquisition device. The vertical acceleration is recorded using the smartphone accelerometer through the free Phyphox application,[19] selecting the "Acceleration (without g)" experiment and a sampling rate of 100 Hz. This measurement removes the gravitational contribution from the recorded acceleration, so the oscillatory signal is approximately centered around zero without requiring any additional constant-offset correction.

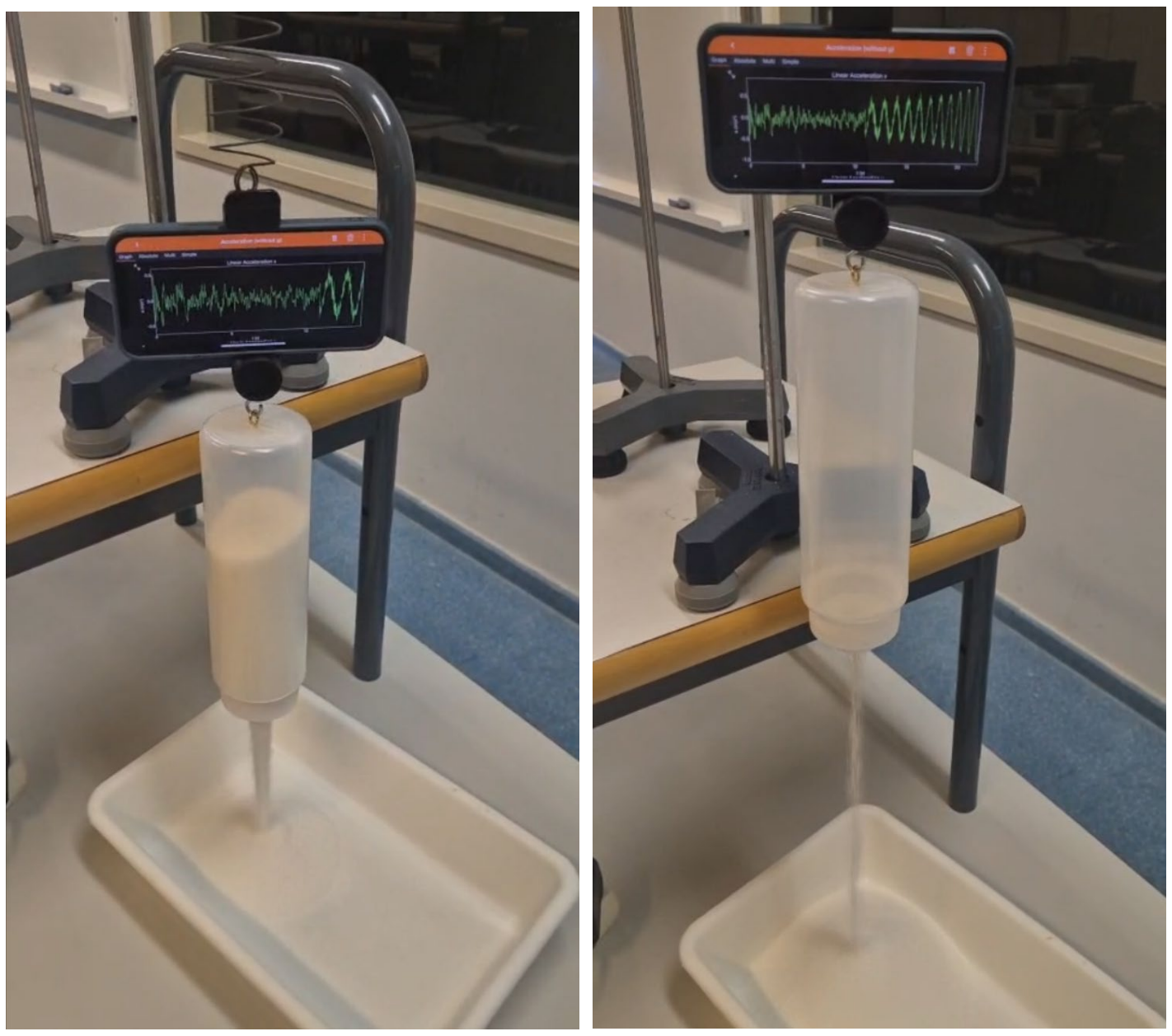

**Fig. 2.** Experimental setup. A smartphone mounted on a spring records the oscillatory motion while a bottle filled with granulated sugar continuously discharges through a 19.55(5) mm outlet, producing an approximately constant mass-loss rate. The images correspond to two frames extracted from the experiment performed with this outlet diameter: an intermediate stage of the discharge process (left) and a late stage near complete emptying of the bottle (right). The acceleration traces displayed on the smartphone illustrate the increase in both oscillation frequency and acceleration amplitude as the oscillating mass decreases. Videos corresponding to the two outlet diameters investigated, 11.60(5) mm and 19.55(5) mm, are provided as Supplemental Material.

A cylindrical plastic bottle is attached to the lower end of the phone holder by means of a small hook. The bottle is filled with granulated sugar, which acts as the variable-mass component of the oscillator. Two outlet diameters were investigated, 11.60(5) mm and 19.55(5) mm, producing two different mass-loss rates. The sugar flows continuously into a collection tray placed beneath the apparatus. As the sugar leaves the bottle, the total oscillating mass decreases while the system performs vertical oscillations. The acceleration signal is recorded continuously from the moment the bottle is released until the discharge process is completed.

The mass of the smartphone together with the mounting system was 227(1) g. The empty bottle had a mass of 56(1) g, whereas the mass of the bottle filled with granulated sugar was 670(1) g. The oscillations were produced using a spring with a spring constant of $k$ = 22.5(1) N/m and a mass of 57(1) g. The initial effective oscillating mass $m_0$ is then obtained by adding the masses of the smartphone, mounting system, and filled bottle to one-third of the spring mass, yielding $m_0 = 916(2)$ g. Neglecting the effective contribution of the spring mass would lead to an overestimation of the oscillation frequency of approximately 1.1% at the beginning of the experiment, increasing to approximately 3.3% as the bottle approaches the empty condition. Thus, including the effective spring mass is particularly important towards the end of the discharge process.

Fig. 2 presents two representative frames extracted from the experiment corresponding to the 19.55(5) mm outlet diameter. The complete video corresponding to this experiment, together with a second video recorded using the 11.60(5) mm outlet diameter, is included as Supplemental Material. The left image corresponds to an intermediate stage of the discharge process, while the right image was recorded near the end of the experiment. Even from these snapshots, the acceleration traces displayed on the smartphone screen reveal qualitative features predicted by the theoretical model. As the mass decreases, successive maxima and minima occur closer together in time, indicating an increase in the oscillation frequency. At the same time, the acceleration amplitude gradually increases, in agreement with the asymptotic behavior derived in Sec. 2.

A key parameter of the theoretical model is the mass-loss rate $\alpha$, which characterizes the decrease of the oscillating mass with time. To determine this parameter independently of the oscillation measurements, the bottle was filled with granulated sugar and suspended above a collection container placed on a digital balance. As the sugar flowed from the bottle, the accumulated mass collected in the container was recorded as a function of time. This procedure is similar to that employed in previous studies of variable-mass systems to characterize the discharge process and verify the assumption of a constant mass-loss rate.[13] This study also verified that the mass-loss rate is the same for a stationary and an oscillating bottle.

The collected mass was measured throughout the discharge process and subsequently analyzed using linear regression. Since the theoretical model assumes a constant mass-loss rate, the accumulated mass is expected to increase linearly with time. The procedure was repeated for both outlet diameters, providing two independent determinations of the mass-loss rate. The resulting measurements therefore provide an independent determination of the parameter $\alpha$ and allow the validity of the linear mass-loss assumption to be assessed experimentally.

The analysis of the discharge measurements is presented in Sec. 4.A. There, the values of $\alpha$ obtained for the two outlet diameters from the mass-versus-time data are compared with those extracted from fits of the theoretical expressions developed in Sec. 2 to the acceleration measurements. Agreement between both determinations provides a direct validation of the proposed variable-mass oscillator model.

# 4. Results and Discussion

### A. Determination of the mass-loss rate

A key assumption of the theoretical model developed in Sec. 2 is that the mass of the oscillator decreases linearly with time according to $m(t) = m_0 - \alpha t$. To verify this assumption and determine the mass-loss rate independently of the oscillation measurements, the discharge process was characterized experimentally using the procedure described in Sec. 3.

The bottle was filled with sugar and suspended above a collection container placed on a digital balance. As the sugar flowed through the outlet hole, the mass accumulated in the container was recorded as a function of time. Measurements were performed for two outlet diameters, 11.60(5) mm and 19.55(5) mm, producing two different mass-loss rates. The resulting mass-versus-time data are shown in Fig. 3. Different symbols are used to distinguish the two outlet diameters, while the solid lines represent linear least-squares fits to the corresponding datasets. The excellent linearity of the measurements confirms that the discharge process can be accurately described by a constant mass-loss rate throughout the duration of the experiment.

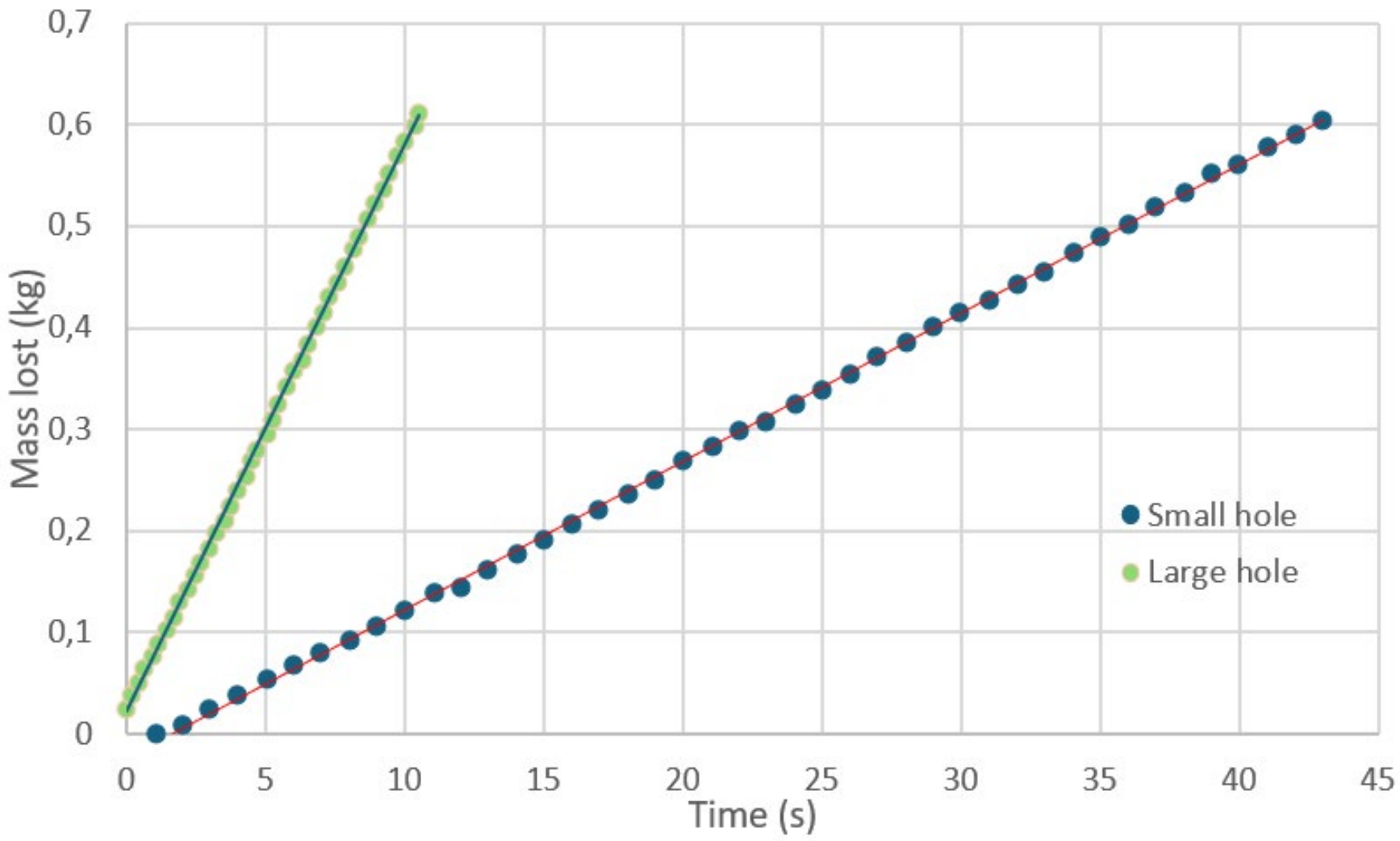

**Fig. 3.** Mass collected in the container as a function of time for outlet diameters of 11.60(5) mm and 19.55(5) mm. Different symbols represent the experimental measurements, while the solid lines correspond to linear least-squares fits. The excellent linearity confirms that the mass-loss rate remained approximately constant throughout the experiment. According to the variable-mass model, the slope of the fitted line provides an independent determination of the mass-loss rate α.

The accumulated mass in the collection container can be described by the linear relation $m_c(t) = \alpha t + m_c(0)$, where $m_c(t)$ is the mass collected at time $t$, $m_c(0)$ is the initial reading of the balance, and the slope of the fitted line corresponds directly to the mass-loss rate $\alpha$.

The slope of the fitted line therefore provides an independent determination of the parameter $\alpha$. For the 11.60(5) mm outlet, the linear fit yielded $\alpha_1 = 0.01460(4)$ kg/s whereas for the 19.55(5) mm outlet the corresponding value was $\alpha_2 = 0.05572(18)$ kg/s. The corresponding coefficient of determination was $R_1^2 = 0.9997$ and $R_2^2 = 0.9996$, both very close to unity. These results indicate that the collected mass increased linearly throughout the discharge process for both outlet diameters and therefore support the assumption of a constant mass-loss rate adopted in the theoretical model. As expected, the larger outlet diameter produced a significantly greater value of $\alpha$.

The values of $\alpha$ obtained from the discharge measurements will be used as a reference in the following sections. In Sec. 4.B, they will be compared with the values obtained from nonlinear fits of the theoretical acceleration model, whereas in Sec. 4.C they will be compared with the values extracted from the predicted linear dependence of $T^2$ on time. Agreement among these independent determinations provides a direct assessment of the validity of the proposed variable-mass oscillator model under two different mass-loss conditions.

**B. Characterization of the variable-mass oscillator**

The acceleration data recorded by the smartphone accelerometer were analyzed using the asymptotic solution derived in Sec. 2.A. Particular care was taken when initiating the oscillations in order to minimize unwanted lateral motions and other spurious perturbations. Several experimental runs were performed, and only those exhibiting a clean predominantly vertical oscillatory motion were selected for further analysis. The selection was based on the absence of appreciable lateral motion or other disturbances that could compromise the comparison with the one-dimensional theoretical model.

The experimental runs analyzed in this section correspond to the two discharge processes characterized in Sec. 4.A, using outlet diameters of 11.60(5) mm and 19.55(5) mm. In both experiments the same spring was employed, with spring constant $k$ = 22.5(1) N/m, allowing the influence of the mass-loss rate to be investigated while keeping the restoring force unchanged.

The experiments correspond to the same configuration described in Sec. 3 and illustrated in Fig. 2. The videos provided as Supplemental Material offer a direct visualization of the progressive increase in oscillation frequency as the oscillating mass decreases, together with the gradual growth of the oscillation amplitude. Both effects are clearly visible in the experimental recordings and constitute qualitative evidence supporting the theoretical predictions developed in Sec. 2.

The recorded acceleration data were fitted using the asymptotic expression given by Eq. (11). The fitting procedure treated the amplitude constant $C$, the phase constant $\phi$, and the mass-loss rate $\alpha$ as adjustable parameters, while the spring constant $k$ and the initial mass $m_0$ were fixed at their independently measured values. The optimal parameter values were obtained by minimizing the difference between the theoretical predictions and the experimental acceleration data using a nonlinear least-squares procedure. The uncertainties reported for the fitted mass-loss rates also account for the propagation of the uncertainties in the independently measured spring constant and initial effective oscillating mass, which contribute approximately 0.5% to the relative uncertainty.

Fig. 4 shows the experimental acceleration data together with the best-fit theoretical curve for the smaller outlet diameter of 11.60(5) mm. Despite the long duration of the discharge process, extending over more than 40 s, the model reproduces accurately both the gradual increase in oscillation frequency and the evolution of the acceleration amplitude. The fitted parameters obtained from this adjustment are summarized in Table I. The fitted value of the mass-loss rate is in good quantitative agreement with the independently measured value obtained from the discharge measurements presented in Sec. 4.A.

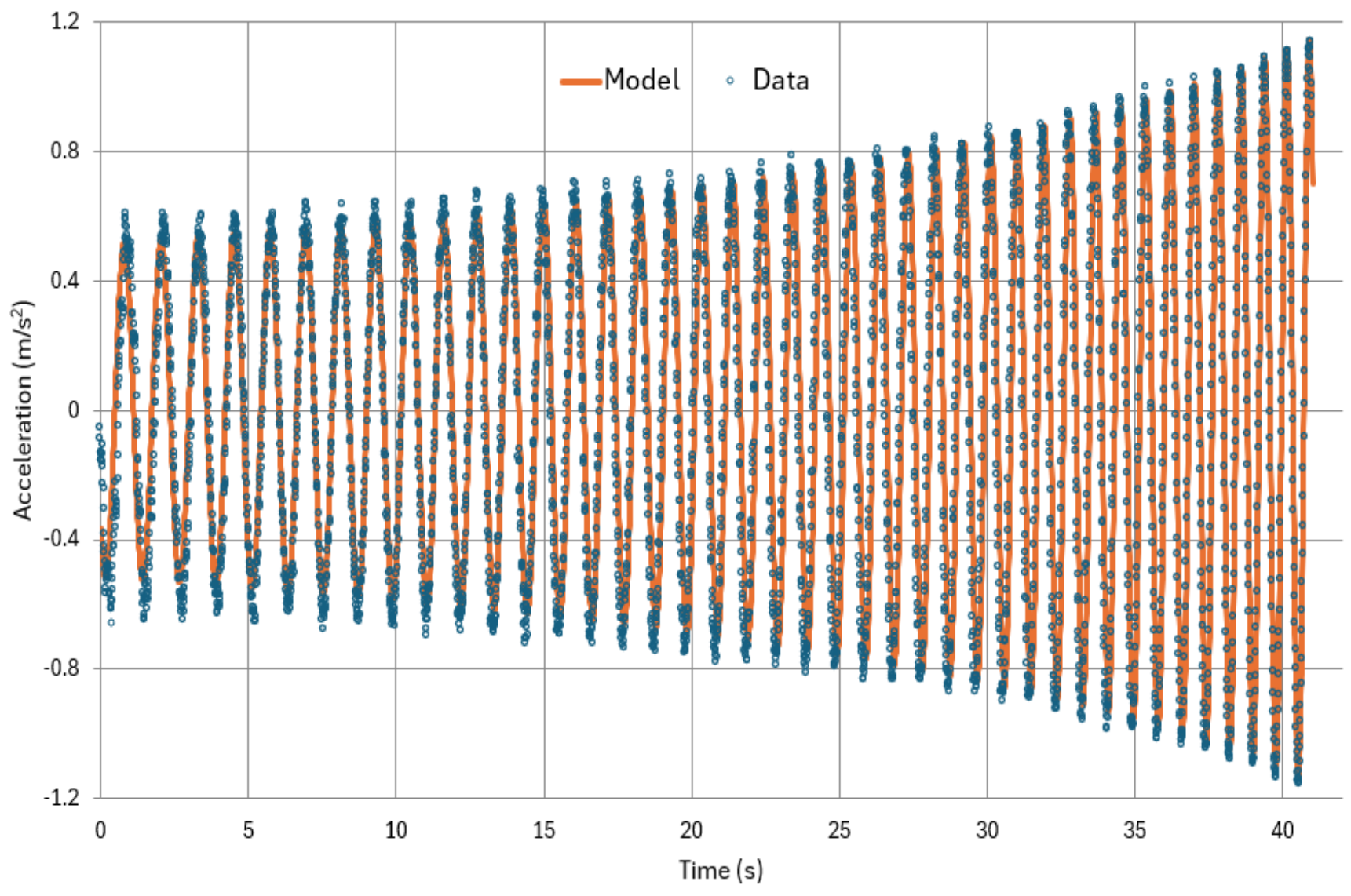

**Fig. 4.** Experimental acceleration data and best-fit theoretical curve obtained from Eq. (11) for the outlet diameter of 11.60(5) mm. The fitted model accurately reproduces the observed evolution of both oscillation frequency and acceleration amplitude during the discharge process.

The corresponding results for the larger outlet diameter of 19.55(5) mm are shown in Fig. 5. In this case the larger opening produces a significantly greater mass-loss rate, resulting in a more rapid evolution of the oscillatory motion. Nevertheless, the theoretical model continues to provide an excellent description of the experimental data. The fitted parameters are listed in Table I and again show good agreement with the independently measured mass-loss rate obtained from the discharge experiment.

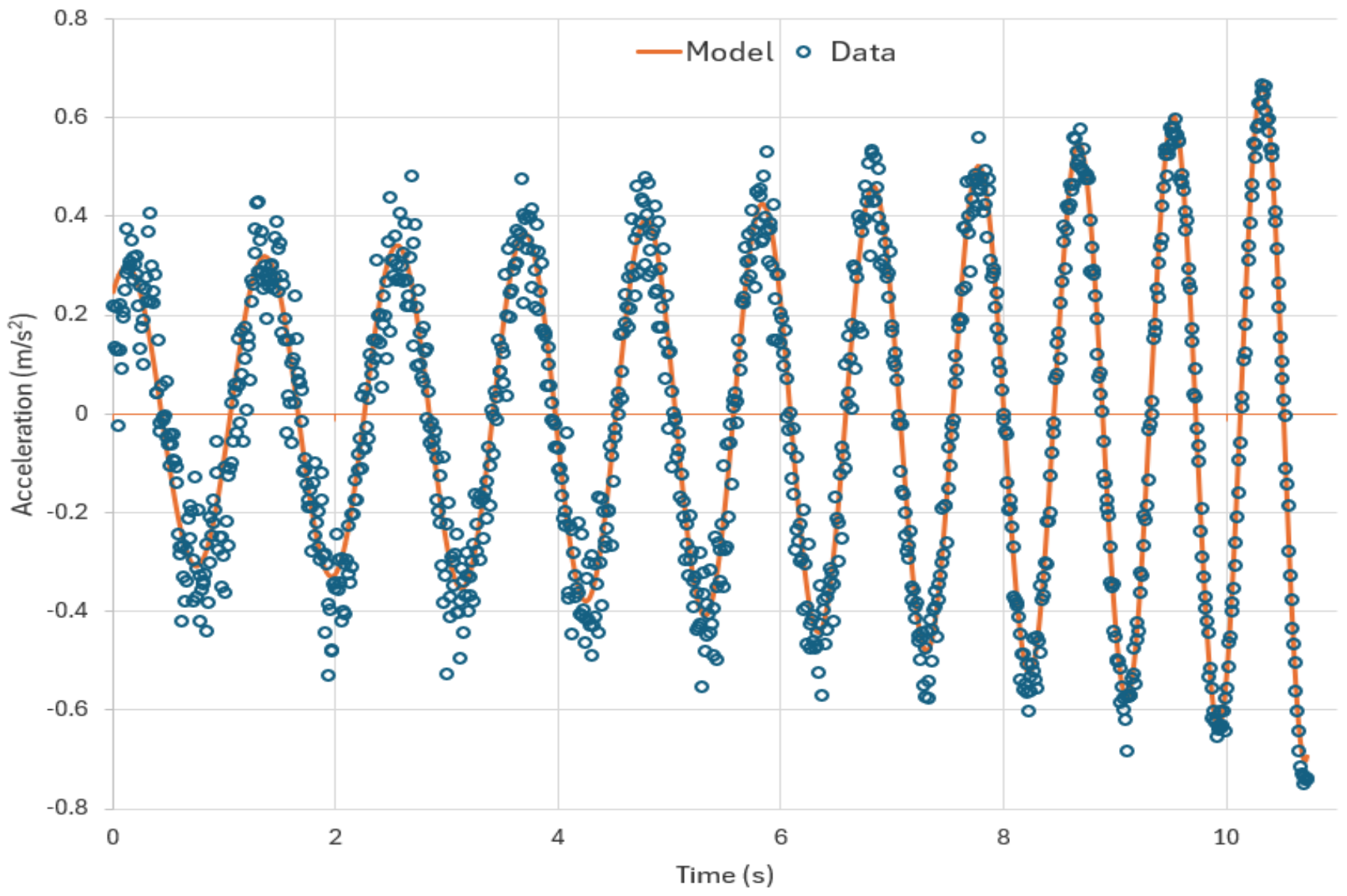

**Fig. 5.** Experimental acceleration data and best-fit theoretical curve obtained from Eq. (11) for the outlet diameter of 19.55(5) mm. The same spring was used as in Fig. 4, allowing the effect of the larger mass-loss rate to be isolated.

The parameters obtained from the two fitting procedures are summarized in Table I. The amplitude constant $C$ and phase constant $\phi$ were treated as free fitting parameters but are not reported, since they depend on the initial conditions of each experimental run and are not directly relevant to the physical interpretation of the results. Table I therefore focuses on the physically significant parameters of the model: the mass-loss rate $\alpha$ and the coefficient of determination $R^2$. The fitted values of $\alpha$ are compared with the independently measured values obtained from the discharge experiments, and the corresponding discrepancies are also reported.

**Table I.** Parameters obtained from fits of the experimental acceleration data using Eq. (11).

| **Outlet diameter (mm)** | 11.60(5) | 19.55(5) |
|---|---|---|
| $\boldsymbol{R^2}$ | 0.952 | 0.957 |
| $\boldsymbol{\alpha_{fit}\ (kg/s)}$ | 0.01465(11) | 0.0555(4) |
| $\boldsymbol{\alpha_{direct}\ (kg/s)}$ | 0.01460(4) | 0.05572(18) |
| **Discrepancy (%)** | 0.3 | 0.4 |

As shown in Table I, the fitted values of the mass-loss rate are in good agreement with those obtained independently from the discharge measurements. The relative discrepancies are only 0.3% and 0.4% for the smaller and larger outlets, respectively, with differences consistent with the reported uncertainties. This agreement is particularly significant because the two outlet diameters produce mass-loss rates differing by nearly a factor of four. Possible sources of systematic uncertainty include small deviations from a perfectly constant mass-loss rate, residual lateral motion, and other experimental effects not explicitly accounted for in the idealized model. Although these contributions have not been quantified separately, they may influence the fitted parameters. The small discrepancies observed support both the validity of the discharge characterization presented in Sec. 4.A and the applicability of the theoretical model to different mass-loss conditions.

The increase in oscillation frequency and acceleration amplitude observed in the experimental videos is quantitatively reproduced by the fitted theoretical curves in both cases. In particular, the model successfully captures the progressive shortening of the oscillation period associated with the continuous reduction of the oscillating mass, as well as the evolution of the acceleration amplitude throughout the discharge process.

The good agreement obtained for two substantially different mass-loss rates using the same spring provides a stringent test of the theoretical framework developed in Sec. 2. The model accurately reproduces the experimental dynamics and recovers values of the mass-loss rate that are consistent with those obtained independently from the discharge measurements, providing strong support for the proposed description of the variable-mass oscillator.

### C. Time-dependent oscillation frequency

As discussed in Sec. 2.B, the asymptotic solution predicts that the square of the oscillation period, $T^2$, should vary linearly with time according to Eq. (14). This prediction provides an independent experimental test of the variable-mass model that does not require fitting the complete acceleration signal and therefore offers a complementary validation of the theoretical framework.

To test this prediction, the oscillation periods were extracted directly from the experimental acceleration data shown in Figs. 4 and 5. Because the oscillation frequency changes continuously during the discharge process, a local estimate of the period was obtained for each extremum of the signal. For a given maximum, the two nearest minima were identified and the corresponding time intervals were measured. Since each interval represents approximately half a period, a local estimate of the oscillation period was obtained by averaging twice these two intervals. An analogous procedure was applied to each minimum using the two nearest maxima. In this way, a sequence of local periods was determined throughout each experiment. The square of the period, $T^2$, was then calculated for each measurement. A total of 82 extrema were analyzed for the smaller outlet diameter and 21 for the larger outlet. The smaller number of extrema in the latter case reflects the shorter discharge duration, while the finite sampling interval of 0.01 s limits the precision with which the extrema can be located.

The resulting values of $T^2$ are shown in Figs. 6 and 7 for the outlet diameters of 11.60(5) mm and 19.55(5) mm, respectively, together with linear least-squares fits. In both cases, the experimental data exhibit the approximately linear behavior predicted by Eq. (14). Linear least-squares fits were therefore performed to determine the slopes of the corresponding straight lines. Since the spring constant is known independently from the characterization of the spring, the slope of the fitted line can be used to determine the mass-loss rate $\alpha$, while the initial mass $m_0$ was determined by substituting the y-intercept value into Eq. (14). The resulting values are summarized in Table II and compared with those obtained from the discharge measurements presented in Sec. 4.A.

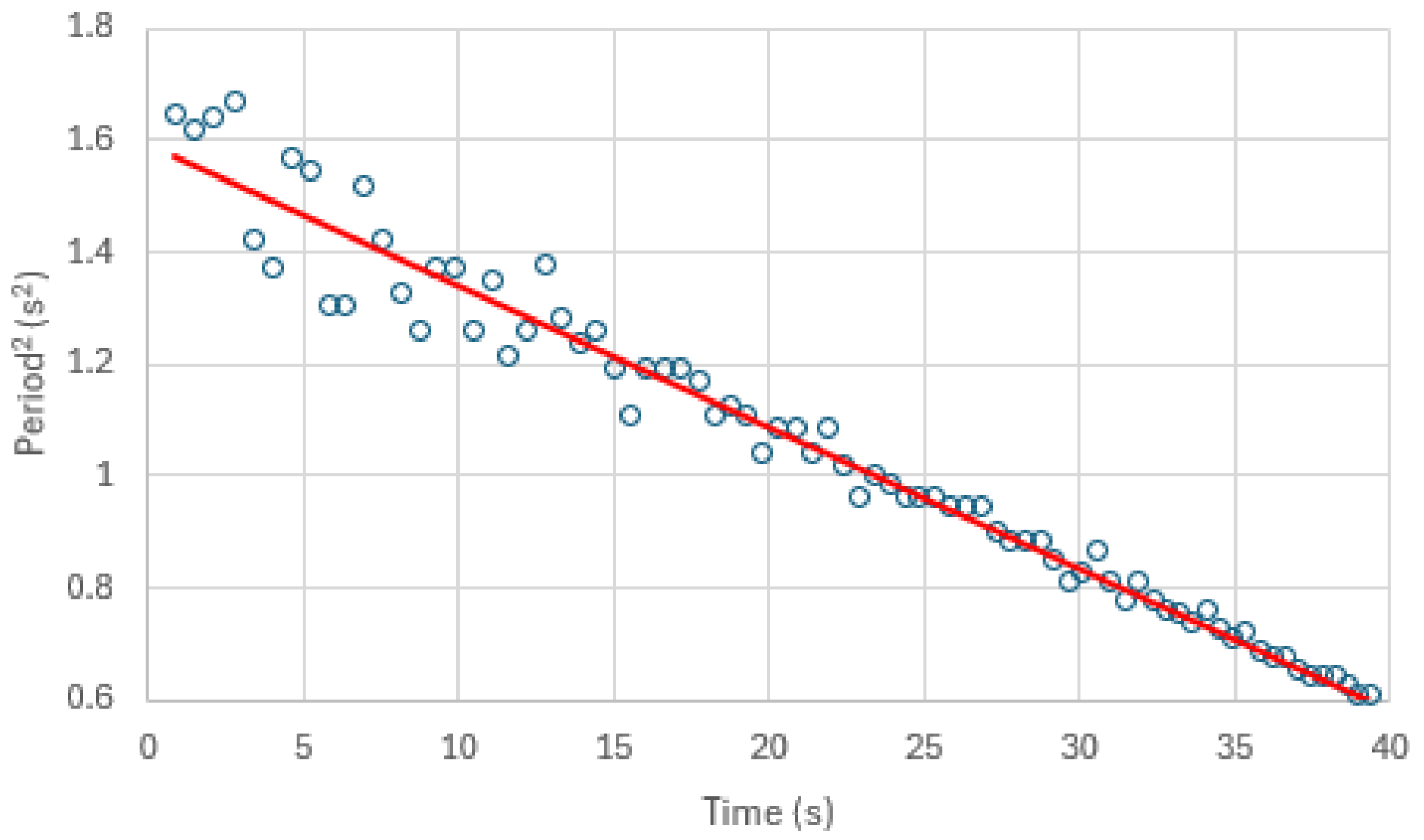

**Fig. 6.** Experimental values of $T^2$ as a function of time for the outlet diameter of 11.60(5) mm. The solid line represents a linear least-squares fit. According to Eq. (14), the slope of the fitted line is proportional to the mass-loss rate $\alpha$, providing an independent determination of this parameter.

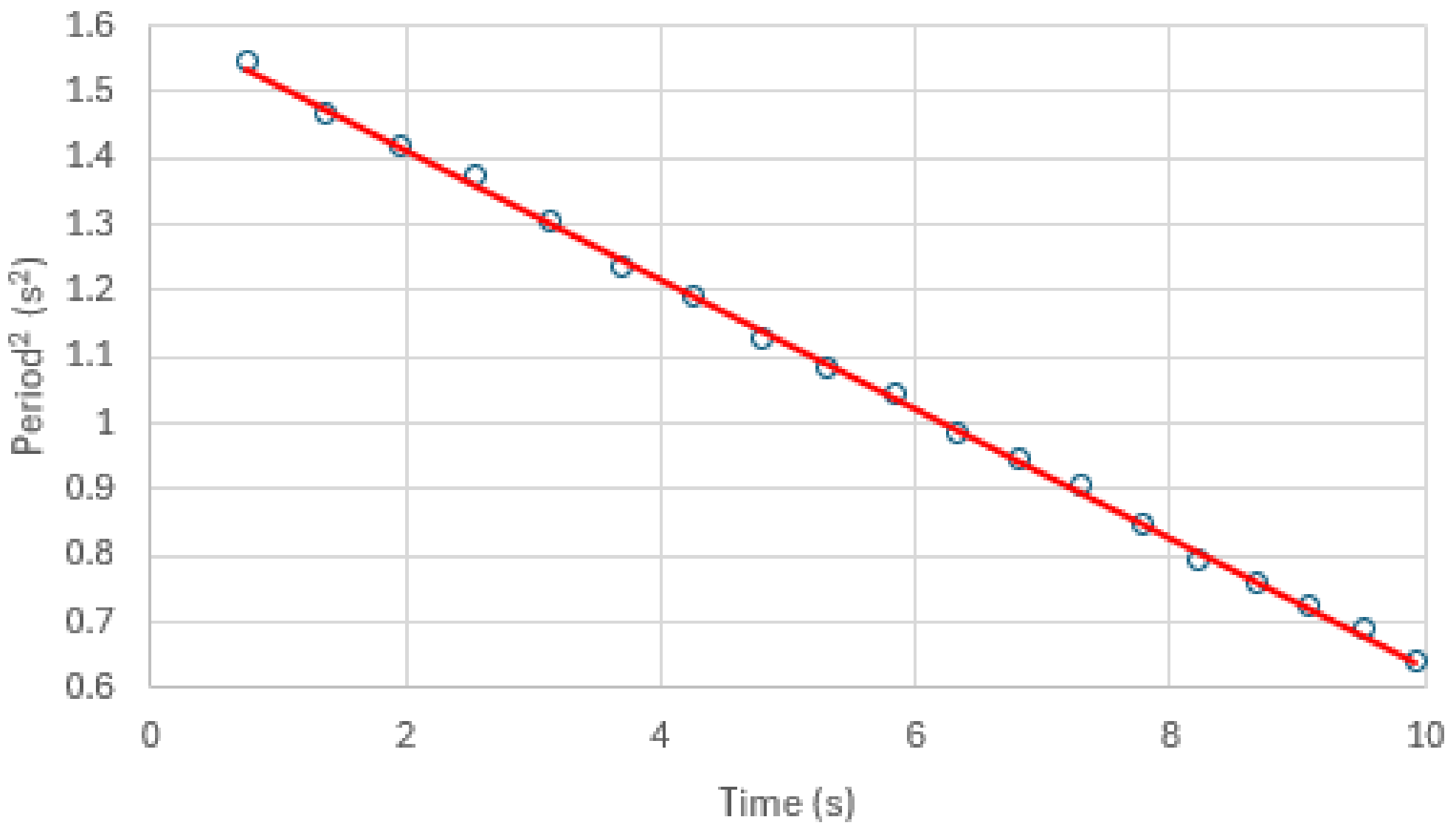

**Fig. 7.** Experimental values of $T^2$ as a function of time for the outlet diameter of 19.55(5) mm. The solid line represents a linear least-squares fit. According to Eq. (14), the slope of the fitted line is proportional to the mass-loss rate $\alpha$, providing an independent determination of this parameter.

**Table II.** Parameters obtained from the linear fits of $T^2$ versus time using Eq. (14).

| **Outlet diameter (mm)** | 11.60(5) | 19.55(5) |
|---|---|---|
| $R^2$ | 0.9563 | 0.999 |
| $\alpha_{fit}$ **(kg/s)** | 0.01439(28) | 0.0556(4) |
| $\alpha_{direct}$ **(kg/s)** | 0.01460(4) | 0.05572(18) |
| **Discrepancy (%)** | 1.4 | 0.2 |
| $m_0$ **(g) from fit** | 909(7) | 916(2) |
| $m_0$ **(g) measured** | 916(2) | 916(2) |
| **Discrepancy (%)** | 0.8 | 0 |

As shown in Table II, the values of $\alpha$ obtained from the period analysis are in good agreement with those determined independently from the discharge measurements, with the observed differences remaining well within the experimental uncertainties. The predicted linear dependence of $T^2$ on time is clearly observed in both experiments. The initial mass values obtained from the fit also match well to the value obtained by weighing.

The agreement among the values of α obtained from the discharge measurements, the fits to the acceleration data, and the period analysis provides strong support for the validity of the theoretical model. In particular, the results confirm the prediction that the oscillation period decreases as the oscillating mass decreases, and that the evolution of $T^2$ is quantitatively described by Eq. (14).

An important advantage of this method is that it only requires identifying successive extrema in the oscillation signal. Consequently, it provides a simple experimental route to determine the mass-loss rate without fitting the complete theoretical expression for the acceleration, making it particularly suitable for instructional laboratory settings.

From an educational perspective, the proposed experiment could be incorporated into an advanced undergraduate mechanics laboratory through a sequence of complementary activities. Students could first determine the mass-loss rate by measuring the accumulated mass as a function of time, then record the oscillatory motion using a smartphone accelerometer and analyze the evolution of the oscillation period. Finally, they could compare the experimental results with the theoretical predictions and discuss the assumptions and limitations of the variable-mass model, particularly the treatment of momentum transport and the validity of the asymptotic approximation. This sequence would provide a practical connection between Newton's laws for open systems, differential equations, and experimental data analysis.

## 5. Conclusions

We have presented and experimentally validated a variable-mass harmonic oscillator consisting of a spring–mass system whose mass decreases continuously through the controlled discharge of granular material. The motion was monitored using the accelerometer of a smartphone, providing a simple and inexpensive method for acquiring high-quality experimental data.

Analytical expressions were derived for the variable-mass oscillator, including an exact solution in terms of Bessel functions and a simplified asymptotic expression. In the asymptotic regime, the theory predicts an increase in oscillation frequency together with a corresponding decrease in the oscillation period as the mass decreases. The theory also predicts an increase in acceleration amplitude during the discharge process. These predictions were confirmed experimentally through direct analysis of the acceleration signal.

The mass-loss rate was determined independently using three different approaches: direct discharge measurements, fitting of the acceleration data to the theoretical model, and analysis of the time dependence of the oscillation period. The good quantitative agreement among these methods provides strong support for the validity of the theoretical description and confirms the assumption of a nearly constant mass-loss rate throughout the discharge process.

The theoretical model successfully reproduced the observed dynamics for two different outlet diameters and corresponding mass-loss rates using the same spring. The fitted values of the mass-loss rate were found to be in good agreement with those obtained independently from the discharge measurements. In addition, the predicted linear dependence of the square of the oscillation period on time was verified experimentally, providing an independent validation of the asymptotic theoretical model.

Beyond its relevance as an example of variable-mass dynamics, the system constitutes an accessible and versatile laboratory activity. The experiment combines analytical modeling, data acquisition using smartphone sensors, nonlinear fitting, and experimental validation of theoretical predictions using inexpensive and readily available equipment. As such, it provides an attractive educational platform for the study of oscillations and variable-mass systems at the undergraduate level.

## References

[1] J. R. Taylor, Classical Mechanics (University Science Books, Sausalito, CA, 2005).

[2] D. Kleppner and R. Kolenkow, An Introduction to Mechanics, 2nd ed. (Cambridge University Press, Cambridge, 2014).

[3] R. Barrio-Perotti, E. Blanco-Marigorta, J. Fernández-Francos, and M. Galdo-Vega, "Theoretical and experimental analysis of the physics of water rockets," Eur. J. Phys. 31, 1131–1147 (2010).

[4] P. Sullivan and B. Chaplin, “A system to change both mass and applied force,” Phys. Teach. 37, 309–311 (1999).

[5] J. Flores, G. Solovey, and S. Gil, “Flow of sand and a variable mass Atwood machine,” Am. J. Phys. 71, 715–720 (2003).

[6] M. G. Calkin and R. H. March, “The dynamics of a falling chain: I,” Am. J. Phys. 57, 154–157 (1989).

[7] C. A. de Sousa, P. M. Gordo, and P. Costa, “Falling chains as variable-mass systems: Theoretical model and experimental analysis,” Eur. J. Phys. 33, 1007–1020 (2012).

[8] M. Denny, “Balloon and chain: An instructive variable-mass system,” Eur. J. Phys. 42, 025013 (2021).

[9] M. Gürgöze and A. Altınkaynak, “Variable-mass dynamics made explicit: a rotating falling-chain test case,” Eur. J. Phys. 47, 035002 (2026).

[10] D. Prato and R. Gleiser, “Another look at the uniform rope sliding over the edge of a smooth table,” Am. J. Phys. 50, 536–539 (1982).

[11] C. A. de Sousa and V. H. Rodrigues, “Mass redistribution in variable mass systems,” Eur. J. Phys. 25, 41–49 (2004).

[12] K. Nakayama, "Remarks on Newton's Second Law for Variable Mass Systems," Eur. J. Phys. 39, 055002 (2018).

[13] J. Flores, G. Solovey, and S. Gil, “Variable mass oscillator,” Am. J. Phys. 71, 721–725 (2003).

[14] H. Rodrigues, N. Panza, D. Portes Jr., and A. Soares, “A model of oscillator with variable mass,” Rev. Mex. Fis. E 60, 31–38 (2014).

[15] M. T. Caccamo and S. Magazù, “Variable mass pendulum behaviour processed by wavelet analysis,” Eur. J. Phys. 38, 015804 (2017).

[16] Á. Suárez, D. Baccino, M. Monteiro, and A. C. Martí, “Normal coordinates in a system of coupled oscillators and influence of the masses of the springs,” Eur. J. Phys. 42, 015003 (2021).

[17] M. Monteiro and A. C. Martí, “Resource Letter MDS-1: Mobile Devices and Sensors for Physics Teaching,” Am. J. Phys. 90, 328–343 (2022).

[18] J. C. Castro-Palacio, L. Velázquez-Abad, F. Giménez, and J. A. Monsoriu, “Using a Mobile Phone Acceleration Sensor in Physics Experiments on Free and Damped Harmonic Oscillations,” Am. J. Phys. 81, 472–475 (2013).

[19] S. Staacks, S. Hütz, H. Heinke, and C. Stampfer, “Advanced tools for smartphone-based experiments: phyphox,” Phys. Educ. 53, 045009 (2018).